\documentclass[conference]{IEEEtran}
\IEEEoverridecommandlockouts
\usepackage{cite}
\usepackage{amsmath,amssymb,amsfonts}
\usepackage{algorithmic}
\usepackage{graphicx}
\usepackage{textcomp}
\usepackage{xcolor}

\definecolor{darkgreen}{rgb}{0.0, 0.5, 0.0}

\usepackage{array}
\usepackage{listings}
\usepackage{xcolor}

\definecolor{eclipseStrings}{RGB}{42,0.0,255}
\definecolor{eclipseKeywords}{RGB}{127,0,85}
\colorlet{numb}{magenta!60!black}

\newcolumntype{C}[1]{>{\centering\arraybackslash}p{#1}}

\def\BibTeX{{\rm B\kern-.05em{\sc i\kern-.025em b}\kern-.08em
    T\kern-.1667em\lower.7ex\hbox{E}\kern-.125emX}}
\begin{document}

\lstset{ %
  language=Python,
  basicstyle=\ttfamily\scriptsize,
  keywordstyle=\color{blue}\bfseries,
  stringstyle=\color{red},
  commentstyle=\color{darkgreen},
  stepnumber=1,
  numbersep=5pt,
  backgroundcolor=\color{white},
  showspaces=false,
  showstringspaces=false,
  showtabs=false,
  frame=single,
  rulecolor=\color{black},
  tabsize=2,
  captionpos=b,
  breaklines=true,
  breakatwhitespace=false,
  linewidth=0.485\textwidth,
  title=\lstname
}

\renewcommand{\lstlistingname}{Example}

\title{Qiskit HumanEval: An Evaluation Benchmark For Quantum Code Generative Models}

\author{
\IEEEauthorblockN{
Sanjay Vishwakarma\IEEEauthorrefmark{3},
Francis Harkins\IEEEauthorrefmark{3}, 
Siddharth Golecha\IEEEauthorrefmark{4},
Vishal Sharathchandra Bajpe\IEEEauthorrefmark{3},\\
Nicolas Dupuis\IEEEauthorrefmark{1}, 
Luca Buratti\IEEEauthorrefmark{2}, 
David Kremer\IEEEauthorrefmark{3}, 
Ismael Faro\IEEEauthorrefmark{3}, 
Ruchir Puri\IEEEauthorrefmark{1} 
and Juan Cruz-Benito\IEEEauthorrefmark{3}}
\IEEEauthorblockA{\IEEEauthorrefmark{1}IBM Research, Yorktown Heights, NY, USA}
\IEEEauthorblockA{\IEEEauthorrefmark{2}IBM Research, Zurich, Rüschlikon, Switzerland}
\IEEEauthorblockA{\IEEEauthorrefmark{3}IBM Quantum, Yorktown Heights, NY, USA}
\IEEEauthorblockA{\IEEEauthorrefmark{4}IBM Quantum, Gurugram, Haryana, India}}

\maketitle

\begin{abstract}
Quantum programs are typically developed using quantum Software Development Kits (SDKs). The rapid advancement of quantum computing necessitates new tools to streamline this development process, and one such tool could be Generative Artificial intelligence (GenAI). In this study, we introduce and use the Qiskit HumanEval dataset, a hand-curated collection of tasks designed to benchmark the ability of Large Language Models (LLMs) to produce quantum code using Qiskit – a quantum SDK. This dataset consists of more than 100 quantum computing tasks, each accompanied by a prompt, a canonical solution, a comprehensive test case, and a difficulty scale to evaluate the correctness of the generated solutions. We systematically assess the performance of a set of LLMs against the Qiskit HumanEval dataset’s tasks and focus on the models ability in producing executable quantum code. 
Our findings not only demonstrate the feasibility of using LLMs for generating quantum code but also establish a new benchmark for ongoing advancements in the field and encourage further exploration and development of GenAI-driven tools for quantum code generation. 

\end{abstract}

\begin{IEEEkeywords}
Qiskit, Large Language Models, Evaluation benchmarks, Qiskit HumanEval, HumanEval
\end{IEEEkeywords}

\section{Introduction}

Quantum computing holds the promise of significantly advancing our computational capabilities with its potential to offer considerable speed-ups over classical computing in certain classes of problems. However, the problem of creating efficient quantum code remains a challenging task, one requiring expertise and specialization in both quantum information and software engineering skills. To overcome these challenges, there is a growing interest in leveraging Generative Artificial Intelligence (GenAI) technologies to assist in the creation and optimization of quantum code~\cite{10.1007/978-3-319-91152-6_32, qiskit-code-assistant-lad, murillo2024challenges}.

Python has emerged as a leading language in the quantum computing space mainly because of its simplicity, flexibility and broad support by powerful libraries and frameworks~\cite{Fund_2023}. Quantum software frameworks based on Python, such as Qiskit~\cite{Qiskit, javadi2024quantum}, play a crucial role in the computing ecosystem by providing tools that facilitate design, simulation, and execution of quantum workloads. These libraries significantly simplify the implementation of quantum algorithms. However, these libraries have varied feature sets and capabilities, and efficiently coding quantum algorithms remains a challenge.

In this context, GenAI emerges as a promising technology for enhancing the quantum programming process. This approach not only has the potential to speed up and streamline the developmental cycle of quantum software, but also opens the door to potential novel quantum code realizations that might not be intuitive by conventional programming methodologies. AI-mediated quantum code generation can also simplify the quantum software development lifecycle and make quantum computing more accessible to a broader range of users.

Qiskit~\cite{Qiskit} is an open-source quantum computing software development framework provided by IBM. It enables users to program, simulate, and run quantum algorithms on quantum computers. Qiskit offers a comprehensive suite of tools for quantum circuit design, optimization, and execution, making it a versatile platform for both researchers and practitioners in the field of quantum computing. For executing quantum workloads, Qiskit Runtime is a key component within the Qiskit ecosystem specifically designed to enhance the execution efficiency of quantum workloads. It operates both as a service and a programming model that facilitates the building, optimizing, and executing of quantum workloads. This component significantly reduces the overhead associated with setting up and running quantum computations by streamlining the process from development to deployment. It allows users to execute quantum circuits on quantum hardware or simulators with reduced latency and increased throughput. It employs the quantum primitives model for construction and execution of quantum workloads. Quantum primitives provide a simplified and efficient approach to programming execution on quantum systems. These core functions are designed to streamline the use of quantum hardware by abstracting routine operations and optimizing execution strategies. Qiskit Runtime facilitates easier access to quantum hardware by managing and optimizing the orchestration of tasks and resource allocation, thereby simplifying the integration of quantum capabilities into conventional computational workflows. 

This integration of Qiskit and tools like Qiskit Runtime within the dataset ensures that the tasks are grounded in realistic and practical quantum computing scenarios, enhancing the relevance and applicability of research outcomes derived from the dataset.

This paper explores the cutting-edge intersection of GenAI for quantum code generation, specifically in the context of evaluating the capabilities of Large Language Models (LLMs) in generating code for Qiskit SDK. The assessment code presents several challenges~\cite{evalplus, human-eval} for LLMs, including evaluating the quality of the generated code, ensuring syntactical correctness, and verifying the accuracy of the code produced. In this paper, we introduce Qiskit HumanEval (QHE), a variation of the HumanEval benchmark~\cite{human-eval} tailored specifically for quantum computing programming using Qiskit. This benchmark is designed to assess the performance, quality, and knowledge of code generation models in a variety of quantum tasks. These tasks, packaged as an evaluation dataset of prompts, unit tests, and their canonical solutions, are designed to assess the functional correctness of the generated quantum code. To the best of our knowledge, despite some existing publications on using LLMs for quantum computing~\cite{liang2023unleashing, qiskit-code-assistant-lad, murillo2024challenges}, there is no research work evaluating the performance of the quantum code generated by LLMs.

The paper is organized as follows: Section II presents the related work in the area of evaluating code LLMs. Section III presents a summary of the Qiskit HumanEval (QHE) dataset, including descriptions of various prompts, unit tests, and tasks included. Section IV presents the results obtained after evaluating several code LLMs using the QHE evaluation dataset. Section V discusses the subject, presents some conclusions and future directions for this work.

\section{Related work}

There exist various code benchmarks for code LLMs evaluating the performance in code completion~\cite{multipleE,crosscodeeval,repobench},
code repair and editing~\cite{octopack,canitedit}, code reasoning~\cite{cruxeval} or code understanding.
Evaluating the quality of generated code is different than natural language because code can be executed.
For that reason, the most popular benchmarks provide unit-tests that can be run so that models can be evaluated on pass rate. Thus, it is less subjective than natural language benchmarks
which often rely on heuristics such as BLEU or ROUGE. 
HumanEval~\cite{human-eval} and MBPP~\cite{mbpp}, as well as their improved version EvalPlus~\cite{evalplus}, are the most known benchmarks for code generation and program synthesis. They aim to test language comprehension, python coding skills, and algorithms at the level of an average entry-level programmer.
These benchmarks are structured very similarly, with minor differences in how the prompt is formulated for the model: every test case contains an instruction that can be implemented with a stand-alone function, a set of unit tests, and a canonical solution that passes all the unit tests. These benchmarks are manually curated to avoid any possible data leakage. Another relevant benchmark for code generation is DS-1000~\cite{ds-1000} which is aligned with real-world problems in data-science using popular Python SDKs such as NumPy, Scikit-learn, TensorFlow or Pytorch.
Our dataset, which is described in the next section, is close to DS-1000, also providing data science problems on the very popular Python library Qiskit~\cite{Fund_2023}.   

\section{Dataset description}
The Qiskit HumanEval dataset is designed for an automated evaluation system. All tests are written by humans, and are designed to assess the LLM in producing executable and functionally accurate Qiskit code.

It  facilitates a comprehensive evaluation of model proficiency in various aspects of quantum programming, from basic circuit design to the implementation of complex quantum algorithms. The dataset is inspired and based on specifications similar to the HumanEval Dataset~\cite{bigcode-evaluation-harness} but is tailored specifically for the Qiskit SDK and contextualized for quantum computing workloads.

The creation of this dataset is the result of a collaborative effort by a panel of experts in quantum computing and Qiskit. The panel includes Qiskit advocates, members of the Qiskit community, members of IBM Quantum support and documentation, and quantum computing software developers. To ensure accuracy, each team member defined a number of tasks, which were then peer-reviewed by the other panel members. Each task and its components (such as prompts and tests) were reviewed at least once by every member of the panel of experts. Based on the feedback received, tasks were revised and improved until the panel reached a consensus that they were correct in terms of definition, feasibility, clarity, and reproducibility.

\begin{table}
\centering
\caption{Task Categorization in the Qiskit Human Eval Dataset}
\begin{tabular}{p{0.3\textwidth}*{2}{C{0.12\textwidth}}}
 \hline
Category & Number of Tasks \\
 \hline
\textsc{Quantum Circuit Generation} & $28$\ \\
\textsc{Simulation and Execution} & $19$\ \\
\textsc{State Preparation and Analysis} & $7$\ \\
\textsc{Algorithm Implementation} & $14$\ \\
\textsc{Gate Operations and Manipulation} & $17$\ \\
\textsc{Visualization and Post-Processing} & $6$\ \\
\textsc{Advanced Circuit Manipulation} & $8$\ \\
\textsc{Quantum Circuit Serialization} & $2$\ \\
\hline
\end{tabular}
\label{tab:task_categories}
\end{table}

\subsection{Composition and Structure}

The Qiskit HumanEval dataset incudes 101 distinct tasks, organized into eight categories based on the quantum computing functionalities they test. Each test is structured to simulate real-world quantum computing scenarios. These tasks are categorized into several thematic areas, reflecting different facets of quantum computing: This dataset provides a comprehensive array of tasks that encompass a wide range of quantum computing functionalities, from basic circuit construction to advanced algorithm simulations. It aims to challenge and evaluate LLM models on their ability to generate syntactically correct and functionally effective quantum code in diverse cases across quantum computing. All tasks included in the dataset are original and not extracted from any existing tutorials, courses, books, or papers. While the experts responsible for creating the tasks may have been inspired by existing materials, they were required to write each task from scratch. The dataset is not part of the training dataset for the large language models (LLMs) being tested. Additionally, this is reflected in certain tasks where the LLMs are required to work with novel features for which there is even more limited documentation.

The dataset includes an extensive set of problems, each formulated to test different aspects of quantum programming:

\subsubsection{Quantum Circuit Generation} This category includes tasks involving the design and creation of quantum circuits. It emphasizes the use of Qiskit's classes and modules to construct diverse quantum circuits, which serve as fundamental building blocks for executing various quantum algorithms. Each circuit contains a series of quantum gates arranged to perform specific computational tasks. Parametrized and customized instruction sets are also tested here, allowing a variety of circuits to represent different computational needs or hardware constraints, thus enhancing the datasets diversity and applicability. It tests the model's ability to manipulate quantum circuits or construct circuits that meet specific operational criteria.

\subsubsection{Simulation and Execution} Tasks in this category require models to execute quantum circuits on simulators or actual quantum hardware using Qiskits interfaces. This includes running simulations on different quantum simulators (Aer, Statevector), real quantum system execution runs, analyzing results, and evaluating specific properties like expectation values. This process tests the circuits under idealized conditions as well as in environments that simulate real quantum noise, providing insights into their performance and robustness. Additionally, real-world execution on quantum hardware, such as IBM’s quantum processors, is covered to evaluate the practical deployment and scalability of quantum solutions.

\subsubsection{State Preparation and Analysis} This involves tasks focused on preparing specific quantum states and analyzing their properties or dynamics, which is focused on examples based on initialization and detailed examination of quantum states. This category is fundamental for setting the stage for algorithm execution in a quantum workflow. It involves preparing quantum states that are crucial for the proper initiation of computational algorithms and analyzing these states to ensure they meet the requisite criteria for successful algorithm implementation.

\subsubsection{Algorithm Implementation} This category details the implementation of a suite of quantum algorithms represented in the dataset. It includes contextualized quantum circuit generation that set the groundwork for algorithm initialization, and oracle creation procedures for algorithms such as Deutsch-Jozsa, Bernstein-Vazirani, Simon's, and Grover's algorithm, which serve as the foundation for quantum research in various problem domains. Workflows for implementing full routines like the quantum Fourier transform and quantum teleportation protocols are included to demonstrate a few challenging workloads. Each algorithmic implementation is created with its respective theoretical underpinning, practical realization, and experimental considerations.

\subsubsection{Gate Operations and Manipulation} Tasks here involve the manipulation or creation of quantum gates. Here, the focus is on the manipulation, transformation, and strategic use of quantum gates, operators, and layouts within circuits. It includes modifying gates or creating gate-based circuits tailored to meet specific computational or hardware-imposed constraints, which are crucial for optimizing circuit functionality and performance. This tests the LLMs understanding of the fundamental building blocks of quantum circuits and their manipulation to achieve desired quantum states or operations using Qiskit’s classes and modules like the transpiler and quantum information module to achieve the results.

\subsubsection{Visualization and Post-Processing} This category comprises of tasks that require the visualization of quantum operations and the post-processing of the output. It is vital for interpreting the outcomes of quantum operations and effectively communicating these results within the scientific community.

\subsubsection{Advanced Circuit Manipulation} Advanced manipulation techniques such as circuit refinement, optimization, and transpilation are covered under this category. These tasks require complex manipulations of circuits such as optimizing circuit layouts or customizing gate operations for specific hardware constraints to enhance efficiency or compatibility.

\subsubsection{Quantum Circuit Serialization} Covers the serialization and formatting of quantum circuits into standardized descriptions or machine-readable formats. This process facilitates the storage, sharing, and further processing of quantum circuits, ensuring their broader applicability and integration into various quantum computing platforms.

The repository for the paper will contain a dataset that conforms to the following structure format:
\begin{itemize}
  
\item \texttt{task\_id:} Each task within the dataset is assigned a unique identifier formatted as ``\texttt{qiskitCodeAssistant/\{number\}.}'' This identifier is crucial for tracking and referencing tasks within the dataset, facilitating easy access and systematic analysis across different studies or evaluations.

\item \texttt{prompt:} The prompt defines the specific quantum computing problem that the LLM needs to solve by generating appropriate quantum code using Qiskit. It frames the challenge, outlining the requirements and objectives clearly to guide the LLMs code generation process effectively.

\item \texttt{canonical\_solution:} This attribute provides a reference implementation that effectively fulfills the requirements set out in the prompt. Serving as a benchmark, the canonical solution aids researchers and developers in understanding potential approaches to solving the task, offering a baseline against which the LLM-generated solutions can be compared.

\item \texttt{test:} The test component is essential for the automated evaluation of the code generated by the LLM. It includes a Python unit test function that performs a series of checks to ensure that the generated code not only runs correctly but also adheres strictly to the specifications detailed in the prompt. These tests assess various aspects of the output, such as its type, attributes, and functional correctness, providing a robust mechanism for validating the efficacy of the LLM model.

\item \texttt{entry\_point:} The entry point specifies the function within the generated code that the test script will call to begin the evaluation process. This ensures that the evaluation system accurately assesses the correct portion of the LLM's output, focusing on the implementation that is intended to solve the task. It is pivotal in aligning the testing procedure with the intended functionality of the generated code, guaranteeing that the evaluation reflects the true performance of the LLM-generated solution.

\item \texttt{difficulty\_scale:} This attribute categorizes each task by complexity into one of three levels: \texttt{basic}, \texttt{intermediate}, or \texttt{difficult}. The difficulty scale is a critical tool for benchmarking, as it allows researchers to measure the performance of models across tasks of varying complexity. This facilitates a nuanced analysis of the models' capabilities, identifying strengths and weaknesses in handling different levels of quantum programming challenges.

\end{itemize}

\subsection{Diversity and Complexity}

The dataset covers a wide range of workloads in the context of quantum computing from a practitioner's perspective using Qiskit. The complexity of the tasks varies to include entry-level tasks that familiarize models with Qiskit’s syntax and semantics as well as advanced tasks accommodating varying dependencies and workflows. This variety ensures that models are tested across a comprehensive range of scenarios, promoting robustness and adaptability in the models code generation. 

In constructing our dataset, we designed 101 tasks and categorized them based on the anticipated level of difficulty: \texttt{basic}, \texttt{intermediate}, and \texttt{difficult}. The distribution of these tasks is as follows: 54 are designated as \texttt{basic}, 45 as \texttt{intermediate}, and 2 as \texttt{difficult}.

\subsubsection{Basic Level Tasks}
\texttt{Basic} tasks typically involve simple quantum circuits and fundamental Qiskit functionalities. These tasks are straightforward and require minimal language comprehension or logical reasoning. For example, one task (Example 1) involves creating a $\phi^+$ Bell state, simulating it using the Aer simulator, and returning a counts dictionary. This is commonly used as a "hello world" example in quantum computing. The validation process checks whether the result is a dictionary containing "00" and "11" states, with the probability of one state between 0.4 and 0.6. Similarly, in another task (Example 2), the objective is to generate a GHZ State, and, depending on a Boolean parameter, return either a Matplotlib figure of the circuit or the circuit object itself, with validation checks for both outputs.

\subsubsection{Intermediate Level Tasks}
\texttt{Intermediate} tasks require the use of complex quantum algorithms and multiple Qiskit functionalities, enhancing the need for reasoning and problem-solving capabilities. An instance is Example 3, where the task involves designing a CHSH game~\cite{CHSH} circuit that takes bits from Alice and Bob as parameters and returns the measured quantum circuit. The validation includes checks on the number of qubits, circuit depth, and utilized gates. Another task (Example 4) instructs participants to create a pulse schedule using Qiskit Pulse for a specified drive channel, with detailed requirements for pulse duration and amplitude, followed by a delay and a pulse repetition. The verification ensures compliance with the defined attributes.

Other intermediate tasks try to simulate use cases that quantum developers are likely to have, rather than just assessing comprehension and logical reasoning. These tasks focus on creating utility functions to manipulate Qiskit objects. Examples include splitting a circuit at each barrier operation, creating a basic transpiler pass (which users could use as a template), and converting the output from a circuit sampler result to a list of Python \texttt{bool} objects.

\subsubsection{Difficult Level Tasks}
Only two tasks are classified as \texttt{difficult}, signifying their reliance on complex quantum algorithms and a high level of comprehension, reasoning, and problem-solving skills. For instance, Example 5 queries the model with generating an encryption key from the circuit and the sender's basis using the BB84 protocol. This requires an in-depth understanding of the BB84 algorithm and is verified by ensuring the results match the expected outcome under controlled test conditions.

By organizing the tasks into three difficulty levels, we seek to thoroughly assess a wide range of problems solved using Qiskit, from basic Qiskit operations to advanced Quantum Algorithms.

\begin{lstlisting}[caption=Example from the Qiskit HumanEval dataset with ``basic'' difficulty scale.]
from qiskit import QuantumCircuit
from qiskit_aer import Aer
from qiskit.compiler import transpile

def run_bell_state_simulator():
    """
    Define a phi plus bell state using Qiskit,
    simulate it using the Aer simulator,
    and return the counts dictionary.
    """
    bell = QuantumCircuit(2)
    bell.h(0)
    bell.cx(0, 1)
    bell.measure_all()
    backend = Aer.get_backend("aer_simulator")
    result = backend.run(transpile(bell, backend), shots=1000).result()
    return result.get_counts(bell)

def check(candidate):
    result = candidate()
    assert isinstance(result, dict)
    assert result.keys() == {"00", "11"}
    assert 0.4 < (result["00"] / sum(result.values())) < 0.6

run_bell_state_simulator()
\end{lstlisting}

\subsection{Emphasizing Simulated and Real System Runs}
This dataset is designed to also provide a robust framework for assessing the performance of generated quantum code under both simulated conditions and on actual quantum hardware. The dual-environment testing is critical for a comprehensive evaluation of the LLMs code generation capablities in the quantum computing context. The simulation framework used is based on Qiskit Aer and real system runs are based on Qiskit Runtime interfaces to IBM Quantum backends on IBM Quantum.

Simulation plays a critical role in the field of quantum computing, particularly in the developmental stages of quantum algorithms. Within the Qiskit framework, the Aer module provides powerful tools for simulating quantum circuits. Simulators allow for the comprehensive testing of quantum algorithms in a controlled environment. Researchers can explore the theoretical correctness and efficiency of algorithms without the need for actual quantum hardware, which is often less accessible and more costly to operate. Qiskit Aer offers advanced noise models that mimic the behavior of real quantum devices. By running simulations that include realistic noise conditions, developers can predict how algorithms will perform on actual quantum hardware, identifying potential issues and optimizations early in the development process. Simulators help in assessing the resource requirements of quantum algorithms, including the number of qubits needed and the computational complexity. This information is crucial for scaling algorithms and preparing them for deployment on real quantum systems.

While simulations are valuable, testing and validating quantum algorithms on actual quantum hardware is indispensable. The ultimate test of any quantum algorithm's efficacy comes from its performance on real quantum hardware. The Qiskit HumanEval dataset includes runs on actual quantum processors, such as those available through IBM Quantum services. These real system runs are indispensable for validating the feasibility and robustness of quantum code generated from LLM models. Each quantum processor has unique characteristics and imperfections, such as qubit connectivity, gate fidelity, and decoherence times. Running algorithms on actual hardware allows researchers to understand how these factors impact performance and to tailor algorithms to specific hardware constraints. By executing algorithms and code on real quantum systems, researchers can evaluate their practical feasibility and scalability. This real-world testing is crucial for transitioning quantum algorithms from theoretical constructs to practical tools that can solve real-world problems. Real system runs serve as a benchmark for the effectiveness of quantum algorithms. They provide a measure of validation that simulations alone cannot offer, confirming that the algorithms can operate under the constraints and imperfections of actual quantum hardware.

In the context of the Qiskit HumanEval dataset, both simulated and real system runs using Qiskit Runtime are integral for assessing the generative LLM models' ability to produce viable and efficient quantum code. Simulations provide a first layer of validation and optimization, while real system runs offer the ultimate test of the algorithms' applicability and effectiveness in true quantum environments. This dual approach ensures a comprehensive evaluation of quantum code generated and LLMs capablities of generating code for various interfaces, promoting a deeper understanding of LLMs capabilities and limitations for a particular SDK.

\begin{lstlisting}[caption=Example from the Qiskit HumanEval dataset with ``basic'' difficulty scale.]
from qiskit import QuantumCircuit

def create_ghz(drawing=False):
    """
    Generate a QuantumCircuit for a 3 qubit GHZ State and measure it.
    If `drawing` is True, return a Matplotlib drawing of the circuit,
    otherwise return the circuit object.
    """
    ghz = QuantumCircuit(3)
    ghz.h(0)
    ghz.cx(0, 1)
    ghz.cx(0, 2)
    ghz.measure_all()
    if drawing:
        return ghz.draw(output="mpl")
    return ghz

def check(candidate):
    from qiskit.quantum_info import Statevector
    import math
    import matplotlib.pyplot as plt
    
    circuit = create_ghz()
    assert circuit.data[-1].operation.name == "measure"
    circuit.remove_final_measurements()
    ghz_statevector = (
        Statevector.from_label("000") + Statevector.from_label("111")
    ) / math.sqrt(2)
    assert Statevector.from_instruction(circuit).equiv(ghz_statevector)
    
    drawing = create_ghz(drawing=True)
    assert isinstance(drawing, matplotlib.figure.Figure)

create_ghz()
\end{lstlisting}

\begin{lstlisting}[caption=Example from the Qiskit HumanEval dataset with ``intermediate'' difficulty scale.]
from qiskit import QuantumCircuit
from numpy import pi

def chsh_circuit(alice: int, bob: int) -> QuantumCircuit:
    """
    Design the CHSH circuit that takes bits of Alice and Bob as input
    and return the Quantum Circuit after measuring.
    """
    qc = QuantumCircuit(2, 2)
    qc.h(0)
    qc.cx(0, 1)
    qc.barrier()
    if alice == 0:
        qc.ry(0, 0)
    else:
        qc.ry(-pi / 2, 0)
    qc.measure(0, 0)
    if bob == 0:
        qc.ry(-pi / 4, 1)
    else:
        qc.ry(pi / 4, 1)
    qc.measure(1, 1)
    return qc

def check(candidate):
    result = candidate(0, 1)
    assert isinstance(result, QuantumCircuit)
    assert result.num_qubits == 2
    assert result.width() == 4
    assert result.depth() == 4
    assert set({'ry': 2, 'measure': 2, 'h': 1, 'cx': 1}.items()).issubset(set(result.count_ops().items()))

chsh_circuit()
\end{lstlisting}

\subsection{Accessibility and Ethics}
The dataset will be openly available to the research community to promote transparency and reproducibility for this research direction in quantum computing. Care has been taken to exclude any proprietary information or data that could lead to ethical concerns, focusing solely on open source publicly available knowledge and problem scenarios.

\subsection{Compatiblity}
As of the time of writing, the codebase is compatible with Qiskit $\geq1.0$ and is expected to evolve alongside the release of new Qiskit versions that offer enhanced capabilities.

\section{Experiment results}

We tested Qiskit HumanEval on different state-of-the-art open-source Code LLMs. We also introduce in this paper a new Qiskit model.

\begin{lstlisting}[caption=Example from the Qiskit HumanEval dataset with ``intermediate'' difficulty scale.]
from qiskit import pulse
from qiskit_ibm_runtime.fake_provider import FakeBelem
from qiskit.pulse import DriveChannel, Constant, Play, Delay

def pulse_schedule_with_constant_and_delay():
    """
    Using Qiskit Pulse, create a schedule with a constant pulse on drive channel 0,
    featuring a duration of 160 and an amplitude of 0.1 and name this schedule
    'pulse_schedule_with_constant_and_delay'. Use the FakeBelem backend for configuration.
    After creating the pulse, add a delay of 400 to the schedule, then replay the same pulse.
    Return the completed pulse schedule.
    """
    backend = FakeBelem()
    constant_pulse = Constant(duration=160, amp=0.1)
    drive_chan = DriveChannel(0)
    with pulse.build(backend=backend, name='pulse_schedule_with_constant_and_delay') as pulse_sched:
        pulse.play(constant_pulse, drive_chan)
        pulse.delay(400, drive_chan)
        pulse.play(constant_pulse, drive_chan)
    return pulse_sched

def check(candidate):
    pulse_sched = candidate()
    constant_pulses = 0
    delays = 0
    for inst in pulse_sched.instructions:
        if isinstance(inst, Play):
            if isinstance(inst.pulse, Constant):
                constant_pulses += 1
        elif isinstance(inst, Delay):
            delays += 1
    assert constant_pulses == 2, 'Schedule should contain exactly two Constant pulses.'
    assert delays == 1, 'Schedule should contain exactly one Delay instruction.'
    assert pulse_sched.name == 'pulse_schedule_with_constant_and_delay'

pulse_schedule_with_constant_and_delay()
\end{lstlisting}

Our base model is a transformer decoder-only model with 8 Billion parameters using multi-query attention, rotary position embeddings, and $4096$ context window. It was pretrained on $4$ Trillion tokens across $116$ programming languages~\cite{granite-code} and is available on Huggingface (ibm-granite/granite-8b-code-base). 

\begin{lstlisting}[caption=Example from the Qiskit HumanEval dataset with ``difficult'' difficulty scale.]
from qiskit import QuantumCircuit
from qiskit_aer import Aer
from numpy.random import randint

def bb84_circuit_generate_key(senders_basis: [int], circuit: QuantumCircuit) -> str:
    """
    Write the function to generate the key from the circuit and the sender's
    basis generated by the sender using BB84 protocol.
    """
    n = len(senders_basis)
    receivers_basis = randint(2, size=n)
    for i in range(n):
        if receivers_basis[i]:
            circuit.h(i)
    circuit.measure_all()
    key = Aer.get_backend('qasm_simulator').run(circuit.reverse_bits(), shots=1).result().get_counts().most_frequent()
    encryption_key = ''
    for i in range(n):
        if senders_basis[i] == receivers_basis[i]:
            encryption_key += str(key[i])
    return encryption_key

def check(candidate):
    from qiskit.quantum_info import Statevector
    from numpy.random import seed
    seed(12345)
    basis = [1, 0, 0, 1, 1]
    circuit = QuantumCircuit(5)
    circuit.x([3, 4])
    circuit.h([0, 3, 4])
    result = candidate(basis, circuit)
    assert result == "11"

bb84_circuit_generate_key()
\end{lstlisting}
To improve the performance at generating Qiskit, we extended the pre-training with Qiskit python code and jupyter notebooks. Our Qiskit dataset was collected in April 2024, that is after the latest 1.0 Qiskit release. 
 We used only data with the following licenses: Apache 2.0, MIT, the Unlicense, Mulan PSL Version 2, BSD-2, BSD-3, and Creative Commons Attribution 4.0. Whilst there is a large amount of Qiskit in the open-source, much of it is deprecated. To ensure both diversity and quality in our data mixture we decided to keep only data updated after 2023, and oversampled the most recent data from Qiskit~\cite{qk-gh}, Qiskit-Community~\cite{qk-comm-gh}, and Qiskit-Extensions~\cite{qk-ext-gh}, which all contain the highest quality Qiskit code available. After deduplication, and applying various cleaning filters, our Qiskit dataset has about $50$ Million tokens. For extend pre-training, the data is packed in $4096$-long sequences with a separator between each samples. We trained for 3 epochs using a learning rate of $1\times10^{-5}$ decayed with a cosine schedule, and a batch size of $64$.

To improve natural language understanding, we further instruct-tuned the model, following the octopack approach as described in~\cite{octopack}. We mixed chat data from openassistant ($8$k samples) and git commit data from commitpackft ($5$k samples).
We also added Qiskit synthetic data: $2.7$k question/answer pairs generated from tutorials using Mixtral 8x7B Instruct~\cite{mixtral}, and $1$k synthetic 
prompt/code pairs, whose execution accuracy were validated using synthetically-generated unittests.
All the sequences were padded on the left and we used a $2048$ sequence length. We trained on $3$ epochs using a batch size of $32$, and a learning rate of $8\times10^{-6}$ decayed with a cosine schedule.

We present the results on HumanEval (HE) and Qiskit HumanEval (QHE) in Table~\ref{he_qhe_results} for different baselines and our Qiskit model. The baselines include $7$B, $15$B, and $30$+B open-source code LLMs.  
Our Qiskit model \textsc{granite-8b-code-qk} performs better than all the other baselines on QHE. The pass rate jumps $17.8$ points compared with \textsc{granite-8b-code-base} which shows the efficiency of our tuning approach. 
In Table~\ref{qhe_result_difficulty}, we breakout the scores by difficulty scale. Our model is the best at the basic and intermediate levels. No model is currently able to pass either of the $2$ difficult tests.      

\begin{table}
\centering
\caption{HumanEval (HE) and Qiskit-HumanEval (QHE) pass@$1$ computed using greedy decoding.}
\begin{tabular}{p{0.21\textwidth}*{2}{C{0.08\textwidth}}}
 \hline
Model & HE & QHE \\
 \hline
\textsc{CodeLlama-34b-Python-hf} & $\mathbf{52.43}$\% & $26.73$\% \\
\textsc{deepseek-coder-33b-base} & $49.39$\% & $39.6$\% \\
\textsc{starcoder2-15b} & $45.12$\% & $37.62$\% \\
\textsc{codegemma-7b} & $42.68$\% & $24.75$\% \\
\textsc{granite-8b-code-base} & $39.02$\% & $28.71$\% \\
\textsc{granite-8b-code-qk} & $38.41$\% & $\mathbf{46.53}$\%\\
\hline
\end{tabular}
\label{he_qhe_results}
\end{table}

\begin{table}
\centering
\caption{QHE pass counts for the three difficulty levels. There are $54$ basic, $45$ intermediate, and $2$ difficult tests.}
\begin{tabular}{p{0.21\textwidth}*{3}{C{0.05\textwidth}}}
\hline
Model & basic & interm. & difficult \\
\hline
\textsc{CodeLlama-34b-Python-hf} & $19/54$ & $8/45$ & $0/2$\\
\textsc{deepseek-coder-33b-base} & $30/54$ & $10/45$ & $0/2$\\
\textsc{starcoder2-15b} & $26/54$ & $12/45$ & $0/2$\\
\textsc{codegemma-7b} & $20/54$ & $5/45$ & $0/2$\\
\textsc{granite-8b-code-base} & $21/54$ & $8/45$ & $0/2$\\
\textsc{granite-8b-code-qk} & $\mathbf{32/54}$ & $\mathbf{15/45}$ & $0/2$\\
\hline
\end{tabular}
\label{qhe_result_difficulty}
\end{table}

\section{Discussion}

The Qiskit HumanEval dataset represents a significant advancement in the intersection of GenAI and quantum computing. As previously mentioned, this dataset is believed to be the first of its kind, designed specifically to evaluate the performance of code large language models (LLMs) in the context of generating code for quantum computing. As seen throughout this paper, the creation of this dataset has adhered to several strict principles and foundations. Each proposed task is novel, original, and has been peer-reviewed by a panel of experts to ensure the accuracy of the prompts and tests in relation to the intended task. The dataset is not created or designed to cater to or be complacent with models already being trained for Qiskit. As observed in Table \ref{qhe_result_difficulty}, the data shows a similar trend and similar results or challenges across different models when solving problems for various difficulty scales. Although our model \textsc{granite-8b-code-qk} performs better than others, its behavior is aligned or similar to models such as \textsc{deepseek-coder-33b-base}.

Given the dynamic nature of quantum computing and Qiskit, which are subject to continuous advancements, including new algorithms, techniques, and enhancements, we anticipate that the project will evolve accordingly. 
At the time this paper was drafted, the dataset consisted of 101 tests. The complete dataset is scheduled for release in the upcoming months, coinciding with the final publication of this paper. The preliminary public release will offer a set of 150 tests, which align with the descriptions provided in Section 2. The primary objective of this initial release is to evaluate the code Language Models in generating quantum code for various quantum tasks, encompassing different levels of complexity and prerequisite knowledge in both code and quantum computing concepts. In keeping with our commitment to accessibility and community collaboration, the dataset will be made publicly available as open source in Q3 2024. This release will enable users from around the world to contribute to and benefit from the dataset, fostering a diverse and robust quantum computing community. Furthermore, this public availability will enable other code LLMs to assess their performance in the quantum coding domain or even create leaderboards for this subject. To maximize the dataset’s relevance and effectiveness, the QHE team will encourage feedback from all users. This collaborative approach will help in identifying areas for improvement and in ensuring that the dataset meets the diverse needs of the community. Detailed guidelines will be provided for those interested in contributing to the dataset’s development on a public git repository, whether through the addition of new tasks, enhancement of existing ones, or provision of translations and documentation. 

In terms of planned updates, the current dataset is compatible with Qiskit version $\geq1.0$. As the Qiskit SDK continues to evolve, the dataset will need to be updated to ensure compatibility with new Qiskit versions and maintaining its relevance and utility for researchers and practitioners. The dataset will be periodically updated to reflect community user feedback and broaden its scope and utility. These enhancements will be documented in subsequent versions, keeping users informed and equipped with the latest tools.

Considering possible expansions and future iterations of this dataset, there are several opportunities to broaden its scope. Future expansion plans for the dataset may involve incorporating tasks formulated in OpenQASM 3.0~\cite{open-qasm-3}, which would increase its compatibility with various libraries and expand its applicability. Also, currently, the dataset is focused on evaluating code generation. However, it could also be enhanced to include tasks related to code explainability, repair, or translation.

\section*{Acknowledgment}
The authors thank the data and model factory who collected the code data and trained the base model, in particular Mayank Mishra, Rameswar Panda, Gaoyuan Zhang, Matthew Stallone, and Hima Patel. We also thank Atin Sood for helping setting up the Qiskit Code Assistant service, and Xuan Liu for support and management.
\bibliographystyle{IEEEtran}
\bibliography{references}

\begin{thebibliography}{10}
\providecommand{\url}[1]{#1}
\csname url@samestyle\endcsname
\providecommand{\newblock}{\relax}
\providecommand{\bibinfo}[2]{#2}
\providecommand{\BIBentrySTDinterwordspacing}{\spaceskip=0pt\relax}
\providecommand{\BIBentryALTinterwordstretchfactor}{4}
\providecommand{\BIBentryALTinterwordspacing}{\spaceskip=\fontdimen2\font plus
\BIBentryALTinterwordstretchfactor\fontdimen3\font minus
  \fontdimen4\font\relax}
\providecommand{\BIBforeignlanguage}[2]{{%
\expandafter\ifx\csname l@#1\endcsname\relax
\typeout{** WARNING: IEEEtran.bst: No hyphenation pattern has been}%
\typeout{** loaded for the language `#1'. Using the pattern for}%
\typeout{** the default language instead.}%
\else
\language=\csname l@#1\endcsname
\fi
#2}}
\providecommand{\BIBdecl}{\relax}
\BIBdecl

\bibitem{10.1007/978-3-319-91152-6_32}
{J. Cruz-Benito, \textit{et al.}}, ``A deep-learning-based proposal to aid
  users in quantum computing programming,'' in \emph{Learning and Collaboration
  Technologies. Learning and Teaching}, P.~Zaphiris and A.~Ioannou, Eds.\hskip
  1em plus 0.5em minus 0.4em\relax Cham: Springer International Publishing,
  2018, pp. 421--430.

\bibitem{qiskit-code-assistant-lad}
{N. Dupuis, \textit{et al.}}, ``{Qiskit Code Assistant: Training LLMs for
  generating Quantum Computing Code},'' \emph{arXiv preprint arXiv:2405.19495},
  2024.

\bibitem{murillo2024challenges}
{J. M. Murillo, \textit{et al.}}, ``Challenges of quantum software engineering
  for the next decade: The road ahead,'' \emph{arXiv preprint
  arXiv:2404.06825}, 2024.

\bibitem{Fund_2023}
\BIBentryALTinterwordspacing
{Unitary Fund}, ``{2023 Quantum Open Source Survey},'' Dec 2023. [Online].
  Available: \url{https://unitaryfund.github.io/survey-website/}
\BIBentrySTDinterwordspacing

\bibitem{Qiskit}
{Qiskit contributors}, ``Qiskit: An open-source framework for quantum
  computing,'' 2023.

\bibitem{javadi2024quantum}
A.~Javadi-Abhari, M.~Treinish, K.~Krsulich, C.~J. Wood, J.~Lishman, J.~Gacon,
  S.~Martiel, P.~D. Nation, L.~S. Bishop, A.~W. Cross \emph{et~al.}, ``Quantum
  computing with qiskit,'' \emph{arXiv preprint arXiv:2405.08810}, 2024.

\bibitem{evalplus}
{J. Liu, \textit{et al.}}, ``Is your code generated by chatgpt really correct?
  rigorous evaluation of large language models for code generation,'' 2023.

\bibitem{human-eval}
{M. Chen, \textit{et al.}}, ``{Evaluating Large Language Models Trained on
  Code.}'' 2021.

\bibitem{liang2023unleashing}
{Z. Liang, \textit{et al.}}, ``Unleashing the potential of llms for quantum
  computing: A study in quantum architecture design,'' \emph{arXiv preprint
  arXiv:2307.08191}, 2023.

\bibitem{multipleE}
{F. Cassano, \textit{et al.}}, ``Multipl-e: A scalable and extensible approach
  to benchmarking neural code generation,'' 2022.

\bibitem{crosscodeeval}
{Y. Ding, \textit{et al.}}, ``Crosscodeeval: A diverse and multilingual
  benchmark for cross-file code completion,'' 2023.

\bibitem{repobench}
T.~Liu, C.~Xu, and J.~McAuley, ``Repobench: Benchmarking repository-level code
  auto-completion systems,'' 2023.

\bibitem{octopack}
{N. Muennighoff, \textit{et al.}}, ``{OctoPack: Instruction Tuning Code Large
  Language Models},'' 2023.

\bibitem{canitedit}
{F. Cassano, \textit{et al.}}, ``Can it edit? evaluating the ability of large
  language models to follow code editing instructions,'' 2023.

\bibitem{cruxeval}
{A. Gu, B. Rozière, \textit{et al.}}, ``Cruxeval: A benchmark for code
  reasoning, understanding and execution,'' 2024.

\bibitem{mbpp}
{J. Austin, \textit{et al.}}, ``Program synthesis with large language models,''
  2021.

\bibitem{ds-1000}
{Y. Lai, \textit{et al.}}, ``Ds-1000: A natural and reliable benchmark for data
  science code generation,'' 2022.

\bibitem{bigcode-evaluation-harness}
{L. Ben Allal, \textit{et al.}}, ``{A framework for the evaluation of code
  generation models},''
  \url{https://github.com/bigcode-project/bigcode-evaluation-harness}, 2022.

\bibitem{CHSH}
\BIBentryALTinterwordspacing
{J. F. Clauser, \textit{et al.}}, ``Proposed experiment to test local
  hidden-variable theories,'' \emph{Phys. Rev. Lett.}, vol.~23, pp. 880--884,
  Oct 1969. [Online]. Available:
  \url{https://link.aps.org/doi/10.1103/PhysRevLett.23.880}
\BIBentrySTDinterwordspacing

\bibitem{granite-code}
{M. Mishra, \textit{et al.}}, ``Granite code models: A family of open
  foundation models for code intelligence,'' \emph{arXiv:2405.04324}, 2024.

\bibitem{qk-gh}
``Qiskit,'' \url{https://github.com/Qiskit/qiskit}, 2024.

\bibitem{qk-comm-gh}
``{Qiskit Community},'' \url{https://github.com/qiskit-community}, 2024.

\bibitem{qk-ext-gh}
``{Qiskit Extensions},'' \url{https://github.com/Qiskit-Extensions}, 2024.

\bibitem{mixtral}
{A-Q. Jiang, \textit{et al.}}, ``Mixtral of experts,'' \emph{arXiv:2401.04088},
  2024.

\bibitem{open-qasm-3}
\BIBentryALTinterwordspacing
{A. Cross, \textit{et al.}}, ``Openqasm 3: A broader and deeper quantum
  assembly language,'' \emph{ACM Transactions on Quantum Computing}, vol.~3,
  no.~3, sep 2022. [Online]. Available: \url{https://doi.org/10.1145/3505636}
\BIBentrySTDinterwordspacing

\end{thebibliography}
\end{document}